\documentclass[aps,prb,floatfix,amsmath,preprint,eqsecnum,nofootinbib]{revtex4}
\usepackage{graphicx}
\usepackage{dcolumn}
\usepackage{bm}
\usepackage{setspace}   

\newcommand{\beq}{\begin{equation}}
\newcommand{\eeq}{\end{equation}}
\newcommand{\ba}{\begin{array}{ccc}}
\newcommand{\ea}{\end{array}}
\newcommand{\nn}{\nonumber}
 \renewcommand{\d}{\partial}
\def\bea{\begin{eqnarray}}
\def\eea{\end{eqnarray}}

\def\<{\langle}
\def\>{\rangle}
\usepackage{amsmath}
\usepackage{amssymb}
\begin{document}

\title{Edge and impurity response in\\ two-dimensional quantum antiferromagnets}

\author{Max A. Metlitski}
\affiliation{Department of Physics, Harvard University, Cambridge MA
02138}

\author{Subir Sachdev}
\affiliation{Department of Physics, Harvard University, Cambridge MA
02138}

\date{August 4, 2008\\
\vspace{1.6in}}
\begin{abstract}
Motivated by recent Monte-Carlo simulations of H\"oglund and Sandvik (arXiv:0808.0408), we study edge response in square lattice
quantum antiferromagnets. We use the O(3) non-linear $\sigma$-model to compute the decay asymptotics of 
the staggered magnetization, energy density and local magnetic susceptibility away from the edge. We find that the total
edge susceptibility is negative and diverges logarithmically as the temperature $T \to 0$. We confirm
the predictions of the continuum theory by performing a $1/S$ expansion of the microscopic Heisenberg model with the edge. We propose a qualitative
explanation of the edge dimerization seen in Monte-Carlo simulations by a theory of valence-bond-solid correlations in the N\'eel state. We also discuss
the extension of the latter theory to the response of a single non-magnetic impurity, and its connection to the theory of the deconfined critical point.
\end{abstract}

\maketitle

\section{Introduction}

The Heisenberg antiferromagnet on a square lattice is one of the best known model magnetic systems. It has been studied extensively both numerically by quantum Monte-Carlo and analytically by $1/S$ expansion and field-theoretic methods. It is known to have an ordered ground state at zero temperature with the staggered magnetization reduced by quantum fluctuations to $N_b = \langle N \rangle = 0.307$ for the spin $S = 1/2$.\cite{SandvikHeis} 

\begin{figure}[h]
\begin{center}
\includegraphics[width=3.5in]{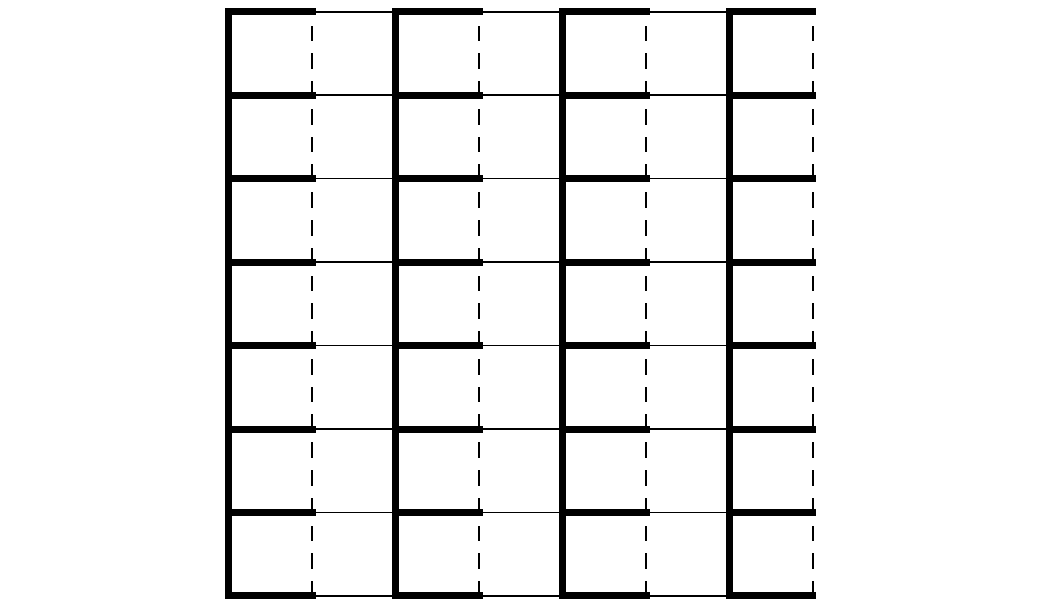}
\caption{A schematic picture of the comb structure in bond strengths observed in Monte-Carlo simulations \cite{Hoglund}, with a free
edge on the left side.}
\label{figcomb}
\end{center}
\end{figure}

Despite many years of study, the simple Heisenberg model does not cease to surprise us. Recent Monte-Carlo simulations\cite{Hoglund} on the $S = 1/2$ model have shown that the edge response in this system is very peculiar. In particular, a negative edge susceptibility is observed at low temperatures. This result is in contrast with an intuitive picture of a ``dangling" edge spin leading to an enhancement in the susceptibility. The simulation of local susceptibility near the edge shows that the negative sign does not come from the edge spins per se, whose susceptibility is, indeed, enhanced, but rather from a tail in the response decaying away from the edge. Another curious effect observed in Ref.~\onlinecite{Hoglund} is the dimerization of bond response near the edge, leading to the appearance of a comb-like structure, as in Fig.~\ref{figcomb}. The tendency to dimerize into singlets near the edge was  argued in Ref.~\onlinecite{Hoglund} to be the source of negative edge susceptibility.

\begin{figure}[h]
\begin{center}
\includegraphics[width=3.5in]{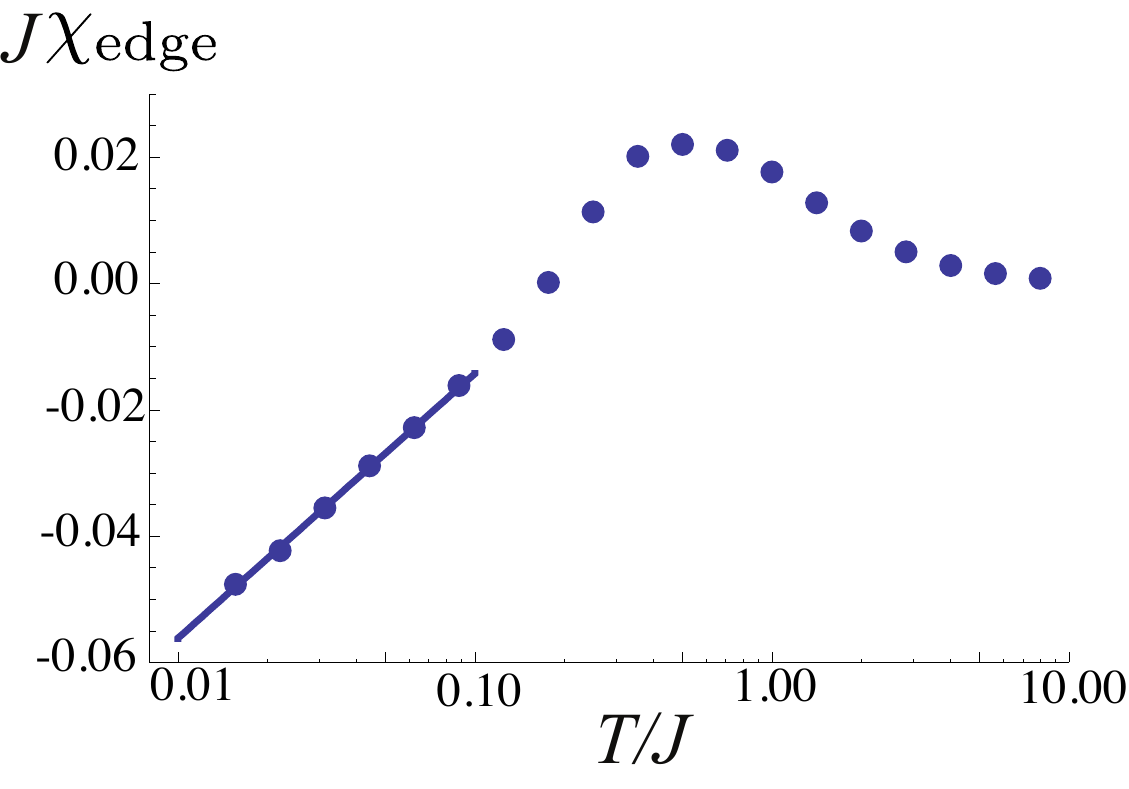}
\caption{Edge susceptibility: Comparison of the Monte Carlo data of Ref.~\onlinecite{Hoglund} (dots) with the best fit line
$J \chi_{\rm edge} = -0.0182 \log(0.219J/T)$ to the low $T$ data.}
\label{fignum}
\end{center}
\end{figure}

In the present paper, we study large-distance asymptotics of the edge response of a square lattice quantum antiferromagnet by means of an effective O(3) $\sigma$-model description. 
This field-theoretic method is an expansion in powers of energy and momentum, with the microscopic physics entering at each order through a finite number of parameters, such as the spin-wave speed $c$, the spin stiffness $\rho_s$ and the value of the staggered moment $N_b$.\footnote{We will use the subscript $b$ from here on to denote bulk properties.} 
The O(3) $\sigma$-model has proved powerful for studying finite temperature/size effects, which typically lead to a crossover into an O(3) model of lower dimension.\cite{Halperin} It turns out to be also useful for studying the edge behaviour, particularly as no new parameters beyond the bulk ones are needed to describe the leading low temperature, large distance asymptotics in the edge response. We concentrate our attention on the staggered moment $\langle N(x) \rangle$, the local energy density $\langle \epsilon(x)\rangle$ and the local magnetic susceptibility $\chi_{\perp}(x)$. We show that at zero temperature these quantities approach their bulk values away from the edge with simple power law forms,
\bea \frac{\langle N(x) \rangle - N_b}{N_b} &=& - \frac{c}{8 \pi \rho_s x}\\
\langle \epsilon(x) \rangle - \epsilon_b &=& \frac{c}{16 \pi x^3}\\
\chi_{\perp}(x) - \chi_{\perp,b} &=& - \frac{1}{8 \pi x c}\label{chiloc}\eea
Integrating eq. (\ref{chiloc}), we conclude that the total edge susceptibility per unit edge length is negative and diverges logarithmically with the system size,
\beq \chi_{\perp,\mathrm{edge}} = - \frac{1}{8 \pi c} \log(L/a)\eeq
We show that at finite temperature the $1/x$ power law in the susceptibility (\ref{chiloc}) is cut-off for distances larger than the thermal wave-length, $x \gtrsim c/T$, leading to the total edge susceptibility,
\beq \chi_{\perp,\mathrm{edge}} = - \frac{1}{8 \pi c} \log(c/T a)\eeq
Such a log divergent susceptibility is indeed seen in the Monte Carlo simulations \cite{Hoglund}. For the co-efficient of the logarithm
in $\chi_{\rm edge} = (2/3) \chi_{\perp,\mathrm{edge}} $,
with $c= 1.69J$, we find $-0.0157/J$, while the Monte Carlo has a best fit value of $-0.0182/J$ (see Fig.~\ref{fignum}). 
This is in reasonable agreement, with the 
difference probably attributable to difficulties in numerically reaching the asymptotic low $T$ limit.

As for the edge comb structure seen in Ref.~\onlinecite{Hoglund}, this is a short distance phenomenon, which cannot be studied within our continuum O(3) $\sigma$-model. In fact, the standard, ``perturbative" treatment of the O(3) model describes only the low-energy excitations which live near the wave-vector $(\pi, \pi)$ and cannot provide any information about valence-bond-solid correlations, which live near $(\pi,0)$ and $(0, \pi)$. Because these correlations are gapped in the antiferromagnet, they must decay exponentially away from the edge, as seen in Monte-Carlo. To capture the short-distance physics, we have performed a $1/S$ expansion of the Heisenberg model on the lattice with an edge. We find the large-distance asymptotics in agreement with the predictions of our continuum theory. However, we don't reproduce the multiple short-distance oscillations of bond energies away from the edge seen by Monte-Carlo. Instead, we find that the bonds touching the edge are stronger than the bulk, while all the subsequent bonds are weaker. We conclude that the edge dimerization is, likely, a {\em non-perturbative effect\/} in $1/S$, which is invisible in the spin-wave expansion. It is remarkable that such non-perturbative effects are present in the simple $S = 1/2$ Heisenberg model, where the $1/S$ expansion yields quantitatively accurate results for many quantities.

In principle, one may be able to explicitly incorporate the non-perturbative physics in the form of hedgehogs into the semi-classical, large $S$ treatment of the Heisenberg model. The hedgehog configurations are relevant for the dimerization physics, as they carry Berry phases,\cite{Haldane} which endow them with non-trivial quantum numbers under the lattice symmetry.\cite{ReadSachdev} However, studying the hedgehog contribution to the edge physics is technically intractable. 

Instead, we pursue a more phenomenological approach, in which we assume that the system possesses a dynamical valence-bond-solid order parameter with a large correlation length. This assumption is justified close to a phase transition into a valence-bond-solid phase, which can be tuned by adding additional frustrating interactions to the Heisenberg model.\cite{SandvikVBS,MelkoKaul} Moreover, even for the pure, nearest neighbour Heisenberg model with $S = 1/2$, it has been argued long ago\cite{ChubukovSachdev} that the quantum fluctuations are strong enough that the system is ``proximate" to a phase transition at which the magnetic order is lost. This proximity is manifested by the existence of an intermediate temperature window, dominated by the quantum critical point (the low temperature physics is dominated by the antiferromagnet, while the high temperature physics is dominated by the non-universal lattice effects). The observation of edge dimerization over more than 5 lattice spacings in the latest Monte Carlo simulations implies that the correlation length of the valence-bond-solid order parameter in the $S  = 1/2$ Heisenberg model is rather large, further supporting the proximity to a phase transition. 


We show that the comb structure of the bond order seen in Monte-Carlo simulations can be qualitatively understood in the quantum critical language. The particular details of the critical theory are not very important for this purpose - the physics can be read off straight-forwardly from the transformation properties of observables under the lattice symmetry.  In particular, we demonstrate that close to the critical point the oscillations of bonds perpendicular to the edge and lines parallel to the edge in the comb can be related to each other. 

In another recent paper with Kaul and Melko \cite{kmms}, we have discussed the response of the valence bond solid order
to a single non-magnetic impurity (such as a Zn site replacing a Cu site). We used there a phenomenological theory 
similar in spirit to that used
here for the edge response. We will review that theory here and also discuss its connection to the impurity response
in the deconfined theory of the N\'eel to valence bond solid transition discussed in Ref.~\onlinecite{max2}. For this single-site impurity 
case, we are able to infer the non-perturbative role of hedgehogs and Berry phases in somewhat greater detail.

This paper is organized as follows. Section \ref{sec:O3} is devoted to the description of the edge in the framework of the O(3) model at zero temperature. In section \ref{sec:O3T} we discuss the crossover of edge susceptibility to finite temperature. In section \ref{sec:largeS} we perform the large $S$ expansion of the Heisenberg model with an edge. In section \ref{sec:comb} we discuss edge dimerization in a quantum antiferromagnet in the proximity to a phase transition into a valence-bond-solid. Finally, in section~\ref{sec:imp}
we discuss the related theory of the response near a non-magnetic impurity.
Some concluding remarks are presented in section \ref{sec:concl}.

\section{Edge response in the O(3) $\sigma$-model}
\subsection{Zero Temperature}
\label{sec:O3}

In this section we discuss the large distance asymptotic behaviour away from the edge of the staggered moment, local uniform susceptibility and the bond energies using the continuum O(3) $\sigma$-model.
The advantage of this approach is that the results obtained are exact, depending only on a few phenomenological parameters, such as spin-wave velocity $c$, spin-stiffness $\rho_s$ and bulk staggered moment $N_b$. These parameters are known from $1/S$-expansion and Monte-Carlo simulations.

The $\sigma$-model action for the local order parameter $\vec{n}$, satisfying $\vec{n}^2 = 1$, is
\beq S = \frac{\rho^0_s}{2} \int d^2 x d \tau \, (\d_{\mu} \vec{n})^2 \label{SO3} \eeq
Here, we've set $c  =1$, we will restore $c$ at the end of the computations. 
Since we are studying the problem with an edge, we also have to consider boundary perturbations. 
The simplest terms allowed by symmetries are,
\beq S_{\mathrm{bound}} = \sum_{\mu} c_{\mu} \int dy d\tau \, (\d_{\mu} \vec{n})^2 \eeq
This term is irrelevant by power counting (the coupling has scaling dimension $-1$), and can be ignored for the leading asymptotic behaviour calculations performed below. Note that the ``lower dimension" surface term $\vec{n} \d_{x} \vec{n}$ vanishes identically due to the constraint $\vec{n}^2 = 1$. The absence of a boundary term, implies that $\vec{n}$ obeys free boundary conditions,
\beq \d_x \vec{n} = 0 \eeq
as can be seen by varying the action (\ref{SO3}) with respect to $\vec{n}$, integrating by parts and requiring that the surface term be zero. 

To set up perturbation theory, we write $\vec{n} = (\vec{\pi}, \sqrt{1- \vec{\pi}^2})$ and expand the action in $\vec{\pi}$, obtaining,
\beq S = \frac{\rho^0_s}{2} \int d^2 x d\tau \, \left( (\d_{\mu} \vec{\pi})^2 + \frac{1}{1-\vec{\pi}^2} (\vec{\pi} \d_{\mu} \vec{\pi})^2\right)\eeq
The second term in brackets above can be expanded as a power series in $\vec{\pi}$ - yielding terms with couplings of scaling dimension $-1$ and lower. These terms again will not influence the leading asymptotic behaviour of observables discussed below. 

We are, thus, left with the free theory for the Goldstone fields $\vec{\pi}$, supplemented by the free boundary condition $\d_x \vec{\pi} = 0$. The propagator with these boundary conditions is,
\bea \langle \pi^a(\vec{x},\tau) \pi^b(\vec{x}',\tau') \rangle &=& \frac{\delta^{ab}}{\rho^0_s}\int \frac{d \omega}{2\pi} \frac{dk_y}{2 \pi} \frac{dk_x}{\pi} \frac{1}{\omega^2 + k^2_x+k^2_y}
e^{i \omega (\tau - \tau')} e^{i k_y (y-y')} \cos(k_x x) \cos(k_x x')\nn\\ &=& \frac{\delta^{ab}}{\rho^0_s}(D(x-x',y-y',\tau-\tau') + D(x+x',y-y',\tau-\tau'))\label{prop}\eea
where $D(x)$ is the standard $3d$ massless propagator,
\beq D(x) = \frac{1}{4 \pi |x|}\eeq

Now, we can calculate the observables. Let's start with the staggered moment $\langle \vec{N}\rangle$. The microscopic $\vec{N}(x)$ is related to the $O(3)$ field $\vec{n}(x)$ via a multiplicative renormalization, $\vec{N}(x) = N_b Z_N \vec{n}(x)$ where $N_b$ is the exact value of the bulk staggered magnetization and $Z_N$ is a formal power series in $\rho^{-1}_s$, adjusted order by order to give $\langle N^3 \rangle = N_b$ in the bulk.

\begin{figure}[t]
\begin{center}
\includegraphics[angle=0,width = 0.7\textwidth]{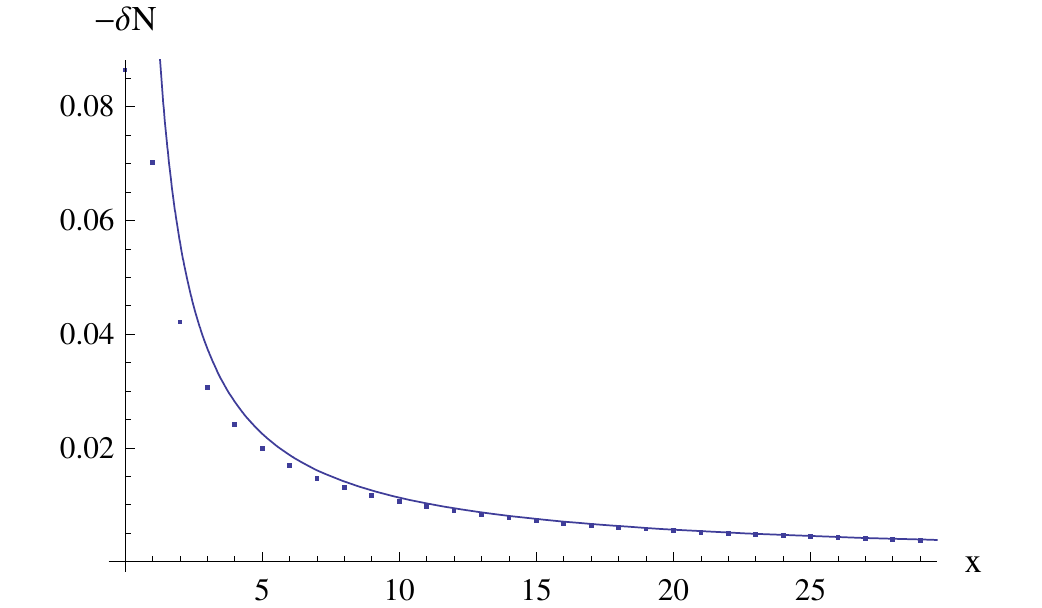}
\caption{Depletion of the staggered moment, $- (\langle \vec{N}(x) \rangle - N_b)$ near the edge. The dotted line is the calculation in the $1/S$ expansion. The solid line is the $O(3)$ $\sigma$-model result for asymptotic behaviour, with phenomenological parameters $\rho_s$, $c$, $N_b$ matched to $1/S$ expansion.}
\label{Nassympt}
\end{center}
\end{figure}

Hence, the staggered moment, to leading order is,
\beq \langle n^3(x)\rangle  = \langle 1 - \frac{\vec{\pi}^2}{2} \rangle = 1 - \frac{1}{\rho^0_s} (D(0) + D(2 x,0,0)) = 1 - \frac{1}{\rho^0_s} (D(0) + \frac{1}{8 \pi x}) \eeq
Thus, as $\lim_{x \to \infty} Z_N \langle n^3(x) \rangle = 1$, and to leading order $\rho^0_s = \rho_s$,
\beq Z_N = 1 + \frac{1}{\rho_s} D(0) = 1 + \frac{1}{\rho_s} \int \frac{d^3k}{(2\pi)^3} \frac{1}{k^2}\eeq
which is the familiar expression known from calculations with no boundary. So,
\beq \langle N^3(x) \rangle = N_b \left(1 - \frac{c}{8 \pi \rho_s x}\right)\label{NO3}\eeq
where we've reinserted the spin-wave velocity $c$. The result (\ref{NO3}) is asymptotically exact and shows suppression of the N\'eel moment near the edge. We can check the result (\ref{NO3}) against the large distance asymptotics of the $1/S$ expansion performed in section \ref{sec:largeS}. The parameters $\rho_s$, $c$ and $N_b$ are known in $1/S$ expansion to be at leading order,
\beq \rho_s = J S^2, \quad c = 2 \sqrt{2} J S a, \quad N_b = S \label{paramS}\eeq
where $a$ is the lattice spacing. Substituting these parameters into (\ref{NO3}) and comparing to our numeric integration results from $1/S$ expansion on the lattice with an edge, we find very reasonable asymptotic agreement (see Fig. \ref{Nassympt}).

Next we consider the uniform transverse susceptibility $\chi_\perp$. Recall, the uniform magnetic field enters (\ref{SO3}) as,
\beq  S_H = \frac{\rho^0_s}{2} \int d^2 x d \tau \, \left((\d_{\tau} n^a - i \epsilon^{a b c} H^b n^c)^2 + (\d_i \vec{n})^2\right)\eeq
The corresponding response function is,
\bea \chi^{ab}(x,x') &=& \frac{\delta^2 \log Z}{\delta H^a(x) \delta H^b(x')} = \rho^0_s (\delta^{ab} - \langle n^a n^b(x)\rangle) \delta^3(x-x') -
(\rho^0_s)^2 \epsilon^{a c d} \epsilon^{b e f} \langle n^c \d_{\tau} n^d (x) n^e \d_{\tau} n^f(x')\rangle\nn\\\label{chiO3exact}\eea
Specializing to the transverse susceptibility, $a,b  =1,2$ and expanding in $\vec{\pi}$,
\bea \chi^{ab}(x,x') 
&\approx& \delta_{\mu \nu} \rho^0_s (\delta^{ab} -  \langle \pi^a(x) \pi^b(x')\rangle) \delta^2(\vec{x}-\vec{x}') \delta(\tau -\tau')\nn\\ &-& (\rho^0_s)^2
\epsilon^{a c} \epsilon^{bd} (\langle \d_{\tau} \pi^c(x) \d_{\tau} \pi^d(x')\rangle
+ (\langle \d_{\tau} \pi^c(x) (\pi^d \vec{\pi} \d_{\tau} \vec{\pi} - \frac{1}{2} \vec{\pi}^2 \d_{\tau} \pi^d)(x')\rangle + (x\leftrightarrow x',
c \leftrightarrow d)))\nn\\
\label{chiO3}\eea
Now, we are actually interested in local response to a static, uniform external field,
\beq \chi^{ab}_{\perp}(x) = \lim_{q_y \to 0} \int dx' dy' d\tau' \chi^{ab}(x,x') e^{-i q_y (y-y')}\eeq
Note that for a finite system size/temperature relevant for Monte-Carlo simulations, at zero external field, there is no distinction between parallel and transverse susceptibility, and we expect,
\beq \chi(x) = \frac{2}{3} \chi_{\perp}(x)\label{perptoiso}\eeq
Since we are working with the static susceptibility, the contribution of the terms in the last two lines of (\ref{chiO3}) is zero, and 
\beq \chi^{ab}_{\perp}(x) =  \rho^0_s (\delta^{ab} - \langle \pi^a(x) \pi^b(x')\rangle) = \rho^0_s \delta^{ab} (1 - \frac{1}{\rho^0_s}
(D(0) + D(2x,0,0)))\eeq
We know that in the bulk, $\chi_{\perp,b} = \lim_{x \to \infty} \chi_{\perp}(x) = \rho_s$ by Lorentz invariance. The bare spin-stiffness $\rho^0_s = \rho_s Z_{\rho}$ where $Z_{\rho}$ is a formal power series in $1/\rho_s$. Thus,
\beq Z_{\rho} = 1 + \frac{1}{\rho_s} D(0) = 1 + \frac{1}{\rho_s} \int \frac{d^3 k}{k^2}\eeq
and we recognize the standard renormalization factor for $\rho_s$. Note that the equality of the first non-trivial terms in $Z_N$ and $Z_{\rho}$ is an accident, which occurs in the O(3) model (for O(N) the coefficients are generally different). Thus,
\beq \chi_{\perp}(x) = \frac{\rho_s}{c^2} - \frac{1}{8 \pi x c}\label{chiT0}\eeq
where we've reinserted $c$. Note that the deviation of $\chi_{\perp}(x)$ from its bulk value is negative, in agreement with Sandvik's simulations. Moreover, the long distance contribution to the total edge susceptibility (per edge length) is given by,
\beq \chi_{\perp,{\mathrm{edge}}} = \int_0^{\infty} dx (\chi_{\perp}(x)-\chi_{\perp,b}) \sim - \frac{1}{8 \pi c} \log (L_x/a)\eeq
At zero temperature, the $\log$ divergence of the long-distance tail will always overpower any short-distance contribution (which can be positive as suggested by the $1/S$ calculation in section \ref{sec:largeS}), leading to a negative total edge susceptibility, as seen by Sandvik. At a finite temperature $T$ (and in the infinite volume limit) the $\log L_x$ divergence will be cut-off at the ``thermal length," $c\, T^{-1}$, leading to 
\beq \chi_{\perp, {\mathrm{edge}}} \sim - \frac{1}{8 \pi c} \log\left(\frac{c}{T a}\right)\eeq
This result will be confirmed by an explicit calculation in the next section.

\begin{figure}[t]
\begin{center}
\includegraphics[angle=0,width = 0.8\textwidth]{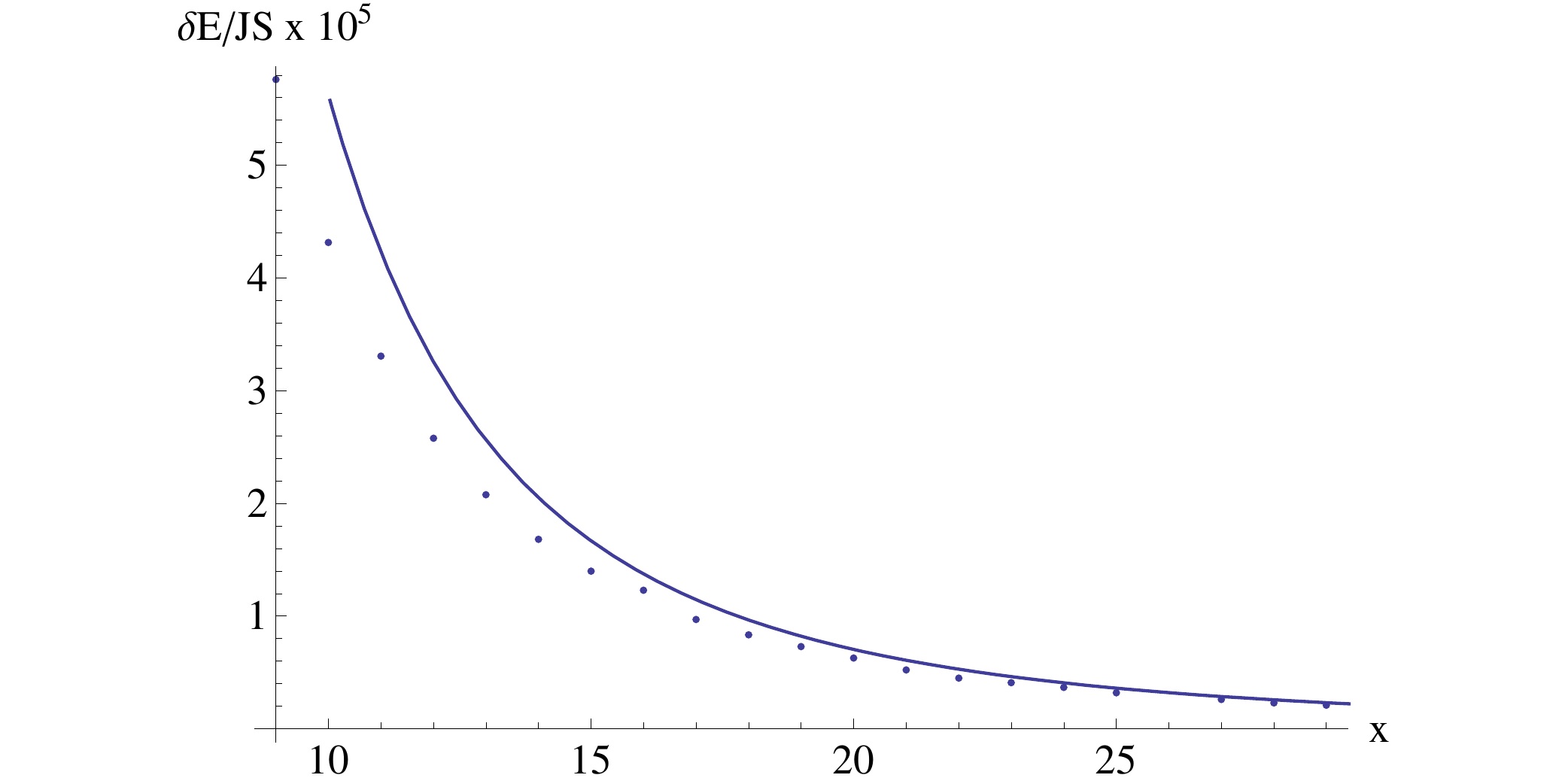}
\caption{Asymptotic increase of local bond energy near the edge. The dotted line is the calculation in the $1/S$ expansion. The solid line is the O(3) $\sigma$-model result for asymptotic behaviour, with phenomenological parameters $\rho_s$, $c$, $N_b$ matched to $1/S$ expansion.}
\label{Bondsassympt}
\end{center}
\end{figure}

Finally, we come to the behaviour of the bond energies. We observe that the sum of bonds energies along the $x$ and $y$ directions is just the local energy density 
\beq \epsilon(x) \sim \frac{J}{a^2} (\vec{S}_i \vec{S}_{i+\hat{x}} + \vec{S}_i \vec{S}_{i+\hat{y}})\eeq For the free field theory describing our Goldstones, in Minkowski space,
\beq \epsilon(x) = \frac{\rho_s}{2} \left((\d_{t} \vec{\pi})^2 + (\d_i \vec{\pi})^2\right) \eeq
Continuing this to Euclidean space,
\beq \epsilon(x) = \frac{\rho_s}{2} \left(-(\d_{\tau} \vec{\pi})^2 + (\d_i \vec{\pi})^2\right) \eeq
Now,
\beq \frac{\rho_s}{2} \langle \d_{\mu} \vec{\pi}(x) \d_{\nu} \vec{\pi}(x) \rangle = \lim_{x \to x'} \frac{{\d}^2}{\d x^{\mu} \d x'^{\nu}}
(D(x-x',y-y',\tau-\tau') + D(x+x',y-y',\tau-\tau'))\label{dmudnu}\eeq
The first term on the righthandside is independent of the distance from the edge and, therefore, we drop it. Noting,
\beq \d_{\mu} \d_{\nu} D(x) = -\frac{1}{4 \pi |x|^3} \left(\delta_{\mu \nu} - 3 \frac{x_{\mu} x_{\nu}}{|x|^2}\right)\eeq 
the second term in (\ref{dmudnu}) yields,
\bea \frac{\rho_s}{2} \langle (\d_{\tau} \vec{\pi})^2(x) \rangle &=& - \d^2_{\tau} D(2x,0,0) = \frac{1}{4 \pi (2x)^3}\\
 \frac{\rho_s}{2} \langle (\d_{x} \vec{\pi})^2(x) \rangle &=&  +\d^2_{x} D(2x,0,0) = \frac{2}{4 \pi (2x)^3}\\
 \frac{\rho_s}{2} \langle (\d_{y} \vec{\pi})^2(x) \rangle &=&  -\d^2_{y} D(2x,0,0) = \frac{1}{4 \pi (2x)^3}\eea
Collecting terms we obtain,
\beq \langle \epsilon(x)\rangle = \frac{c}{16 \pi x^3}\label{T00assympt}\eeq
Note that energy density is enhanced near the edge, corresponding to a decrease of bond strengths, $-\langle \vec{S}_i \vec{S}_j \rangle$.
We can again compare the asymptotically exact expression (\ref{T00assympt}) to the results of the $1/S$ expansion in section \ref{sec:largeS}, by using the parameters (\ref{paramS}). We see from Fig. \ref{Bondsassympt} that the agreement is rather good.

\subsection{\bf Edge susceptibility at finite temperature}
\label{sec:O3T}
To compute the uniform susceptibility at finite temperature $T \ll \rho_s$, we follow the usual strategy of dividing the field $n(\vec{x}, \tau)$ into zero frequency piece, $n(\vec{x})$ and finite frequency modes $\pi_{\alpha}(\vec{x}, \tau)$, 
\beq n^a(\vec{x}, \tau) = \sqrt{1 - \pi_{\alpha}\pi_{\alpha}} n^a(\vec{x}) + \pi_{\alpha}(\vec{x}, \tau) e^a_{\alpha}(\vec{x})\eeq
where $\alpha = 1, 2$ and $\vec{e}_\alpha(\vec{x})$ and $\vec{n}(\vec{x})$ form an orthonormal basis. The strategy is to first integrate over the ``fast" modes $\pi_{\alpha}$ to obtain an effective action for the slow $\vec{n}$ field. Expanding the action in powers of $\pi$ to leading order,
\beq S \approx \frac{\rho^0_s}{2} \int d^2x d \tau \,(\d_{\mu} \pi_{\alpha})^2 + \frac{\rho^0_s}{2} \int d^2 x d \tau \,((\d_i n^a)^2 (1- \vec{\pi}^2) + \d_i e^a_{\alpha} \d_i e^a_{\beta} \pi_{\alpha} \pi_{\beta} + 2 \d_i e^a_{\alpha} e^a_{\beta} \pi_{\alpha} \d_i \pi_{\beta})\eeq
In setting up the perturbation theory in $\pi$ the first term above is treated as the free piece, while the coupling of $\pi$ to the slow fields in the second term is treated as a perturbation. Thus, in a theory with the edge at finite temperature, the bare propagator for the $\pi$ field still satisfies free boundary conditions,
\beq \langle \pi_{\alpha}(\vec{x},\tau) \pi_{\beta}(\vec{x}',\tau')\rangle = \frac{1}{\rho^0_s}\delta_{\alpha \beta} D_n(x,x')\eeq
where,
\beq D_n(x,x') = \hat{D}(x-x',y-y',\tau-\tau) + \hat{D}(x+x',y-y',\tau-\tau')\eeq
with
\beq \hat{D}(\vec{x},\tau) = \frac{1}{\beta} \sum_{\omega_n \neq 0} \int \frac{d^2k}{(2\pi)^2} \frac{1}{k^2 + \omega^2_n} e^{i (\vec{k} \vec{x} + \omega_n \tau)}\eeq
Now, expanding the susceptibility (\ref{chiO3exact}),
\bea \chi^{ab}(x) &=& \rho^0_s (\delta^{ab} - \langle n^a n^b(x)\rangle)- (\rho^0_s)^2 \epsilon^{acd} \epsilon^{bef} \int d^3 x'
 \langle e^c_{\alpha} e^{d}_{\beta}(\vec{x}) e^e_{\gamma} e^{d}_{\delta}(\vec{x}') \pi_{\alpha} \d_{\tau} \pi_{\beta}(x) \pi_{\gamma} \d_{\tau} \pi_{\delta}(x')\rangle\nn\\\label{chiTxxp}\eea
At leading order, we may factorize the correlator of slow $e$ and fast $\pi$ fields in (\ref{chiTxxp}). Moreover, since at finite temperature rotational invariance is restored,
\beq \langle n^a n^b(x) \rangle = \frac{\delta^{ab}}{3} \langle \vec{n}^2(x)\rangle = \frac{\delta^{ab}}{3}\eeq
Hence, the local susceptibility becomes,
\beq	 \chi^{ab}(x) = \frac{2}{3} \rho^0_s \delta^{ab} - (\rho^0_s)^2 \epsilon^{acd} \epsilon^{bef} \int d^3 x'
 \langle e^c_{\alpha} e^{d}_{\beta}(\vec{x}) e^e_{\gamma} e^{d}_{\delta}(\vec{x}')\rangle \langle \pi_{\alpha} \d_{\tau} \pi_{\beta}(x) \pi_{\gamma} \d_{\tau} \pi_{\delta}(x')\rangle \label{convol}\eeq
We see that the susceptibility involves a convolution of correlators of slow and fast fields. Evaluating the correlation function of the fast fields explicitly,
\bea \chi^{ab}(x) &=& \frac{2}{3} \rho^0_s \delta^{ab} -  \epsilon^{acd} \epsilon^{bef} (\delta_{\alpha \gamma} \delta_{\beta \delta}- \delta_{\alpha \delta}\delta_{\beta \gamma}) \int d^3 x' \langle e^c_{\alpha} e^{d}_{\beta}(\vec{x}) e^e_{\gamma} e^{f}_{\delta}(\vec{x}')\rangle (\d_{\tau} D_n(x,x'))^2 \nn\\ \label{convD}\eea
We note,
\bea &&\int d\tau'  (\d_{\tau} D_n(x,x'))^2   =  \frac{1}{\beta} \sum_{\omega_n} \omega^2_n D_n(\vec{x},\vec{x}',\omega_n)^2\nn\\ &=& \frac{1}{\beta} \sum_{\omega_n} \omega^2_n (D(\vec{x}-\vec{x}',\omega_n)^2 + 2 
D(\vec{x}-\vec{x}',\omega_n) D(\vec{x} - R \vec{x}',\omega_n) + D(\vec{x} - R \vec{x}',\omega_n)^2)\nn\\\label{convD2}\eea
where $R$ denotes reflection across the edge at $x = 0$. In the absence of an edge, we can drop the last two terms in (\ref{convD2}). Then we note that the correlation function of $\pi's$ decays exponentially for large distances, hence only $|\vec{x} - \vec{x}'| \lesssim T^{-1}$ contribute to the integral in (\ref{convol}). The slow degrees of freedom $\vec{n}(\vec{x})$ and $\vec{e}_{\alpha}(\vec{x})$  fluctuate only on much larger distances (in fact $T^{-1}$ serves as an effective short-distance cut-off for the slow degrees of freedom), hence we can to leading order set $\vec{x} = \vec{x}'$ in the correlation function of the $e$'s. This leads to a considerable simplification as,
\beq e^a_{\alpha} e^b_{\alpha} = \delta^{ab} - n^a n^b\eeq
and,
\beq (\delta_{\alpha \gamma} \delta_{\beta \delta}- \delta_{\alpha \delta}\delta_{\beta \gamma}) \langle e^c_{\alpha} e^{d}_{\beta}(\vec{x}) e^e_{\gamma} e^{f}_{\delta}(\vec{x})\rangle = \frac{1}{3}(\delta^{ec} \delta^{df} - \delta^{cf}\delta^{de})\eeq
and 
\beq \chi^{ab}(x) = \frac{2}{3} \delta^{ab} \left( \rho^0_s - 2 \int d^3 x' (\d_{\tau} D_n(x,x'))^2 \right)\label{chinoe}\eeq
Now let's introduce the edge back. We wish to compute the deviation of local susceptibility from its bulk value. The major difference from the situation in the bulk is that eq. (\ref{convD2}) no longer depends just on the difference $\vec{x} -\vec{x}'$. For $x T \lesssim 1$,  the integral over $\vec{x}'$ in (\ref{convD}) is saturated with $x' T \lesssim 1$ and hence, we can effectively set $x = x' = 0$, $y = y'$ in the correlation function of the $e$'s and recover the simple form (\ref{chinoe}). However, for $x T \gg 1$, the part of the integral in (\ref{convD}) that represents $\chi(x) - \chi_b$ is no longer saturated at $x' \sim x$. Hence, one really has to compute the correlation function of the slow degrees of freedom. For $T^{-1} \ll x \ll \xi$, we expect this to modify $\chi(x)-\chi_b$ (which, as we shall see, is exponentially suppressed as $e^{- 4 \pi T x}$) by logarithmic corrections. On the other hand, for $x \gtrsim \xi$, we expect additional exponential suppression coming from the slow degrees of freedom. As we shall see, the total edge susceptibility is saturated by $x T \lesssim 1$ and, hence, can be computed directly from (\ref{chinoe}). 

Keeping the above remarks in mind, we obtain
from (\ref{convD2}) and (\ref{chinoe}),
\bea \chi(x) &=& 
\frac{2}{3} \left(\rho^0_s -2 \frac{1}{\beta} \sum_{\omega_n \neq 0}\omega^2_n  \int_{- \infty}^{\infty} dx'\int_{-\infty}^{\infty}dy' (D(\vec{x}-\vec{x}',\omega_n)^2
+ D(\vec{x}-\vec{x}',\omega_n) D(\vec{x}-R \vec{x}',\omega_n) )\right)\nn\\\label{intfull}\eea
The first term under the integral in (\ref{intfull}) is the familiar temperature dependent correction to bulk susceptibility, while the second term represents the edge contribution. Performing the integral over $x'$, 
\beq \chi(x) = \chi_{b}(T) - \frac{4}{3} \frac{1}{\beta} \sum_{\omega_n \neq 0} \frac{d^2k}{(2 \pi)^2} \frac{\omega^2_n}{(k^2+ \omega^2_n)^2} e^{2 i k_x x} \label{chiTx}\eeq
where,
\beq \chi_{b}(T) = \frac{2}{3} \left(\rho^0_s - 2 \frac{1}{\beta} \sum_{\omega_n \neq 0} \int \frac{d^2k}{(2\pi)^2} \frac{\omega^2_n}{(k^2+ \omega^2_n)^2}\right) = \frac{2}{3}\frac{\rho_s}{c^2}(1 + \frac{T}{2 \pi \rho_s})\eeq

Now, we can compute the asymptotics of (\ref{chiTx}). For $x T/c \ll 1$, we can replace the sum over $\omega_n$ by an integral,
\beq \chi(x) \to \chi_{b}(T) - \frac{4}{3} \int \frac{d^3 k}{(2 \pi)^3} \frac{ \omega^2}{(k^2 + \omega^2)^2} e^{2 i k_x x} = 
\chi_{b}(T) - \frac{1}{3} \int \frac{d^2k}{(2\pi)^2} \frac{1}{k} e^{2 i k_x x} = \chi_{b}(T)- \frac{1}{12 \pi x c}\label{chix0assympt}\eeq	
which agrees with our earlier $T = 0$ result (\ref{chiT0}) upon the usual replacement (\ref{perptoiso}). In the opposite limit $x T/c \gg 1$, the sum in (\ref{chiTx}) is going to be dominated by the smallest thermal mass, $\omega_{n = 1}$, and,
\beq \chi(x) \to \chi_{b} - \frac{2}{3} \frac{T}{c^2} \left(\frac{x T}{2c}\right)^\frac12 e^{-4 \pi T x/c}\label{suscdecay}\eeq
As noted earlier, this result will be modified by logarithmic corrections for $x \ll \xi$ and additional exponential suppression for $x \gg \xi$. 
It is also now clear from (\ref{suscdecay}) that the total edge susceptibility is saturated by $x T \lesssim 1$, so that the corrections mentioned above can be ignored for its computation, and we can use eq. (\ref{chiTx}), which obeys the scaling form,
\beq \chi(x) - \chi_{b} = T f_{\chi}(T x)\eeq
Thus, 
\beq \chi_{\mathrm{edge}} = \int_a^{\infty} dx (\chi(x) - \chi_{b}) = \int_{T a}^{\infty} du f(u) \eeq
where $a$ is a short distance cut-off. We observe that the singular behaviour of $\chi_{\mathrm{edge}}$ for $T \to 0$ can be extracted from the short distance asymptotic of $\chi(x)$ (\ref{chix0assympt}). Noting, $f_{\chi}(u) \to - \frac{1}{12 \pi u}$ for $u \to 0$,
\beq \chi_{\mathrm{edge}} \sim - \frac{1}{12 \pi} \int_{T a} \frac{du}{u} = - \frac{1}{12 \pi c} \log\left(\frac{c}{Ta}\right)\eeq
as predicted from $T = 0$ behaviour in the previous section.

\section{Large $S$ expansion of the Heisenberg model with an edge}
\label{sec:largeS}
In this section we perform the large $S$ expansion of the Heisenberg model on a square lattice with an edge. We start with the usual nearest neighbour Hamiltonian,
\beq H = J \sum_{\langle i j \rangle} \vec{S}_i \vec{S}_j\eeq
and use the Holstein-Primakoff representation of spin operators, which at leading order in $1/S$ reads,
\bea S^z_i = S - b^{\dagger}_i b_i, \,\, S^+_i = \sqrt{2S} b_i,\,\, S^-_i = \sqrt{2 S} b^{\dagger}_i, \quad i \in A\\
 S^z_i = -S +c^{\dagger}_i c_i, \,\, S^+_i = \sqrt{2S} c^{\dagger}_i,\,\, S^-_i = \sqrt{2 S} c_i, \quad i \in B\eea
where $A$ and $B$ are the two sublattices. We place the edge at $i_x = 0$. Utilizing the translational invariance along the $y$ direction,
\beq b_{i_x,i_y} = \frac{1}{\sqrt{N_y/2}} \sum_{k_y} b_{i_x, k_y} e^{i k_y i_y},\quad\quad
c_{i_x,i_y} = \frac{1}{\sqrt{N_y/2}} \sum_{k_y} c_{i_x, k_y} e^{i k_y i_y}\eeq
where $-\pi/2 < k_y < \pi/2$ and $N_y$ is the number of sites in the $y$ direction, we obtain the Hamiltonian,
\beq H = 4 S J \sum_k \sum_{i,i'} \left(\begin{array}{c}b_{i,k}\\c^{\dagger}_{i,-k}\end{array}\right)^{\dagger} h_{i i'} \left(\begin{array}{c} b_{i',k}\\c^{\dagger}_{i',-k}\end{array}\right)\eeq
with
\beq
h_{i i'} = \left(\begin{array}{cc} A_{i i'} & B_{i i'}\\B_{i i'} & A_{i i'}\end{array}\right), \quad
 A_{i i'} = \delta_{i i'} (1 - \frac{1}{4} \delta_{i 0}), \quad B_{i i'} = \frac{1}{2} \cos k \delta_{i i'} + \frac{1}{4} (\delta_{i',i+1} + \delta_{i',i-1})\eeq
We perform a Bogoliubov transformation by writing,
\beq \left(\begin{array}{c}b_{i,k}\\c^{\dagger}_{i,-k}\end{array}\right) = \sum_{\lambda > 0} \left(\phi^{+ \lambda}(i) \beta_{\downarrow \lambda, k} + \phi^{-\lambda}(i) \beta^{\dagger}_{\uparrow \lambda, -k}\right)\eeq
where the $\beta$'s obey canonical commutation relations and the two component vectors $\phi^{\lambda}(i) = (u^{\lambda}(i),\, v^{\lambda}(i))$  are eigenstates of $\tau^3 h$,
\bea \tau^3 h \phi^{+ \lambda} &=& \lambda \phi^{+ \lambda} \label{phip}\\
\tau^3 h \phi^{- \lambda} &=& -\lambda \phi^{- \lambda} \eea
Explicitly, $\phi^{- \lambda} = \tau^1 \phi^{+ \lambda}$. We normalize the $\phi$'s as,
\beq \langle \phi^{+ \lambda}| \tau^3| \phi^{+ \lambda'}\rangle = \delta_{\lambda,\lambda'}\eeq
Then, up to a constant,
\beq H = 4 S J \sum_k\sum_{\lambda > 0} \lambda (\beta^{\dagger}_{\uparrow \lambda, k}\beta_{\uparrow \lambda, k} + \beta^{\dagger}_{\downarrow \lambda, k}\beta_{\downarrow \lambda, k})\eeq
The solutions to the eigenvalue problem (\ref{phip}) with positive eigenvalues can be divided into the normalizable and non-normalizable branches. The normalizable branch has dispersion 
\beq \lambda = \frac{1}{\sqrt{2}} |\sin k_y|\eeq
The continuum branch can be parameterized by momentum $0 < k_x < \pi - k_y$ and has dispersion,
\beq \lambda = \sqrt{1 - \frac{1}{4}(\cos k_x + \cos k_y)^2}\eeq
We normalize our continuum solutions to,
\beq \langle \phi(k_x)| \tau^3| \phi(k'_x)\rangle = (2 \pi) \delta(k_x - k'_x)\eeq
Explicit forms of the eigenstates are given in Appendix \ref{app:eigen}. We note that for fixed $k_y \to 0$, the energies of both the normalizable state and the continuum threshold tend to $\frac{1}{\sqrt{2}} |k_y|$, with the splitting between these two energies of order $k^3_y$. This is the reason why the bound state does not show up in the effective low energy O(3) description  - it is treated as being part of the continuum.

Now, we can compute the observables. The staggered magnetization is given by, 
\beq \langle N_j \rangle = S - \langle c^{\dagger}_j c_j \rangle = S - \int_{-\pi/2}^{\pi/2} \frac{dk_y}{\pi} \sum_{\lambda > 0} |v^{\lambda}(j)|^2\eeq
We have evaluated the sum (integral) over the eigenstates numerically - the result is plotted in Fig. \ref{Nassympt}. The staggered moment is depleted near the edge and approaches its bulk value monotonically. If we plug $S  = 1/2$ into our expansion, the staggered moment at the edge is $N_{\mathrm{edge}} = 0.217$ compared to $N_b = 0.303$ in the bulk. As already noted, the long distance asymptotics of the staggered moment are in good agreement with the predictions of the O(3) continuum theory.

\begin{figure}[t]
\begin{center}
\includegraphics[angle=0,width = 0.8\textwidth]{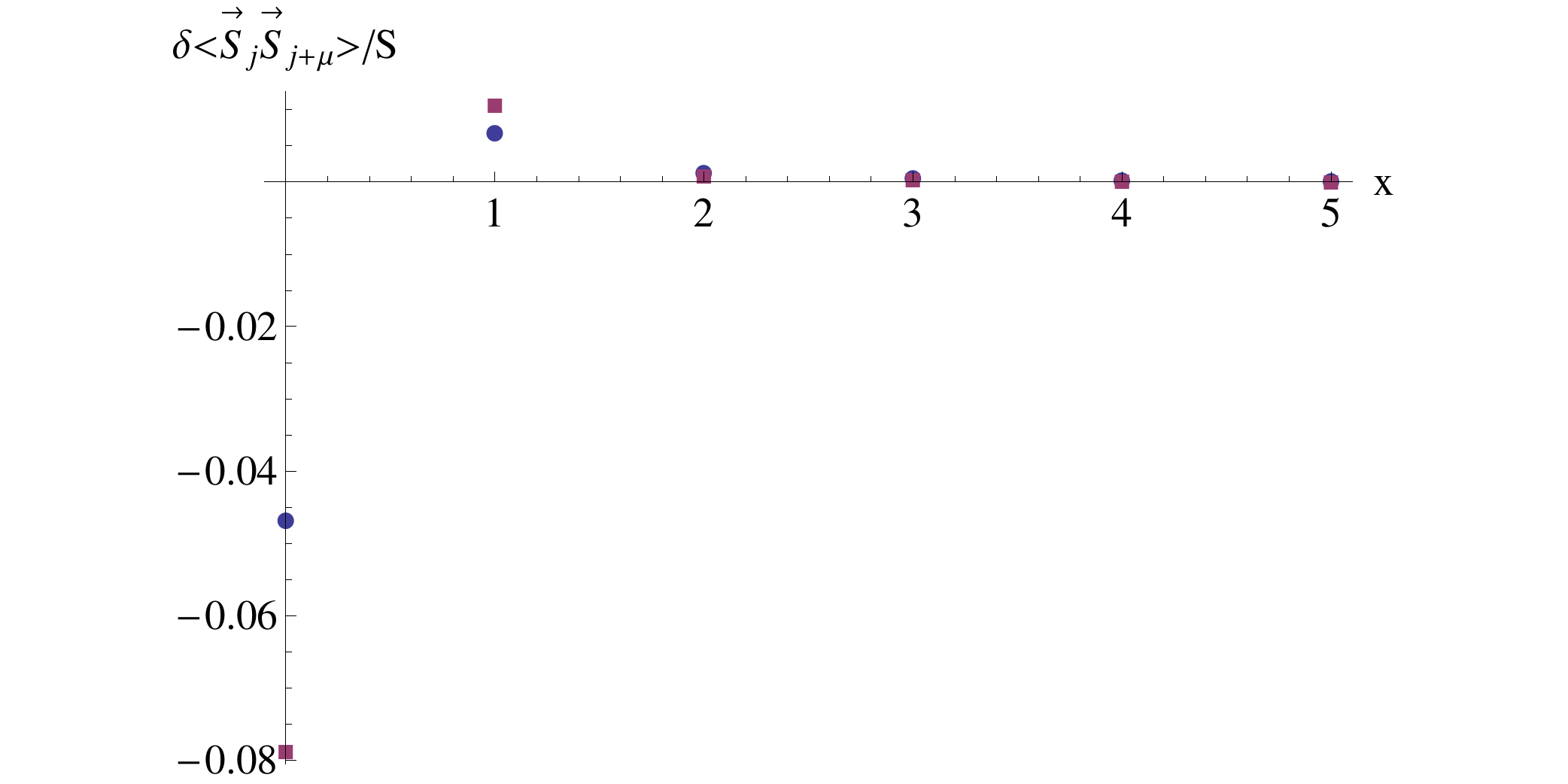}
\caption{Bond strength deviation from bulk value along $x$ (circle) and $y$ (square) directions computed in the $1/S$ expansion.}
\label{Bondsshort}
\end{center}
\end{figure}

Similarly, we can compute the bond energies,
\bea \langle \vec{S}_j \vec{S}_{j + x} \rangle &=& -S^2 + S (\langle b^{\dagger}_j b_j \rangle+ \langle c^{\dagger}_{j+x} c_ {j+x}\rangle + \langle b_j c_{j+x}\rangle + \langle b^{\dagger}_j c^{\dagger}_{j+x}\rangle)\nn\\
&=& -S^2 + S \int_{-\pi/2}^{\pi/2} \frac{dk_y}{\pi} \sum_{\lambda > 0} (|v^{\lambda}(j)|^2 + |v^{\lambda}(j+1)|^2 + v^{\lambda}(j+1)^* u^{\lambda}(j) + u^{\lambda}(j)^* v^{\lambda}(j+1))\nn\\
\langle \vec{S}_j \vec{S}_{j + y} \rangle &=& -S^2 + S ( \langle b^{\dagger}_j b_j \rangle + \langle c^{\dagger}_{j+y} c_{j+y}\rangle  + \langle b_j c_{j+y}\rangle + \langle b^{\dagger}_j c^{\dagger}_{j+y}\rangle)\nn\\ &=& -S^2 + S \int_{-\pi/2}^{\pi/2} \frac{dk_y}{\pi} \sum_{\lambda > 0} (2 |v^{\lambda}(j)|^2 +(u^{\lambda}(j)^* v^{\lambda}(j) + v^{\lambda}(j)^* u^{\lambda}(j)) \cos k_y )\nn\\\eea
The short distance behaviour of the bond energies is shown in Fig. \ref{Bondsshort}. We see that both the perpendicular and parallel bonds touching the edge are stronger than in the bulk ($\langle \vec{S_i} \vec{S_j}\rangle$ is more negative), while all the subsequent bonds are weaker than in the bulk. Substituting $S = 1/2$ into our expansion, we find that at the edge $\langle \vec{S}_j \vec{S}_{j+x} \rangle = -0.352$, $\langle \vec{S}_j \vec{S}_{j+y} \rangle = -0.368$, while in the bulk, $\langle \vec{S}_j \vec{S}_{j+\mu} \rangle = -0.329$. Thus, comparing to the results of quantum Monte Carlo, the $1/S$ expansion reproduces qualitatively the behaviour of the first two rows of bonds away from the edge, but fails to capture the subsequent oscillations in bond strengths on short distances. We expect that these oscillations cannot be seen in the perturbative $1/S$ expansion. In the next section, we will argue that the appearance of such oscillations can be linked to the existence of a competing valence-bond-solid order parameter. As for the long distance asymptotics, we can compare the sum of bond strengths along $x$ and $y$ directions to the local energy density computed in the continuum O(3) model; the two are in good agreement (see Fig. \ref{Bondsassympt}) . 

Now we turn our attention to the local transverse magnetic susceptibility
\beq \chi_{\perp}(j_x) = \frac{1}{2 T N_y} \lim_{q_y \to 0} \sum_{j'_x}\langle S^+(j_x,q_y) S^-(j'_x,-q_y)\rangle\eeq
where 
\beq S^+(j_x,q_y) = \sum_{j_y} S^+(j_x,j_y) e^{-i q_y j_y}\eeq
A finite momentum $\vec{q}$ is needed as a regulator, since we are working in an infinite volume; it is convenient to choose $\vec{q}$ along the $y$ direction. 
At leading order in the $1/S$ expansion,
\bea \chi_{\perp}(j) &=& \frac{1}{2T} S \sum_{j'} \langle (b_{j,q} + c^{\dagger}_{j, -q}) (b^{\dagger}_{j',q} + c_{j',-q})\rangle \\&=& \frac{1}{2 T} S \sum_{j'} \sum_{\lambda > 0} (u^{\lambda}(j,q) + v^{\lambda}(j,q))(u^{\lambda}(j',q) + v^{\lambda}(j',q))^*(1+ 2 n(\lambda))\label{chilat}\eea
where $n(\lambda) = (e^{\lambda/T}-1)^{-1}$ is the bose distribution. As expected, for $q \to 0$, the form-factor in (\ref{chilat}) vanishes upon summing over $j'$, unless $\lambda \to 0$. Thus, we may replace, $n(\lambda) \to T/\lambda$, obtaining,
\beq \chi_{\perp}(j) = S \sum_{j'} \sum_{\lambda > 0} \frac{1}{\lambda} (u^{\lambda}(j,q) + v^{\lambda}(j,q))(u^{\lambda}(j',q) + v^{\lambda}(j',q))^*\eeq
A short calculation then yields,
\beq \chi_{\perp}(j) = \frac{1}{8 J} (1+ (-1)^j (\sqrt{2}+1)^{-(2j + 1)})\label{chiS}\eeq
This result is saturated by normalizable modes and states at the bottom of the continuum band. We see that as $j \to \infty$, the susceptibility approaches its bulk value
$\chi_{\perp,b} = \frac{1}{8 J}$. We can define the edge susceptibility (per unit edge length) as,
\beq \chi_{\perp,\mathrm{edge}} = \sum_j (\chi_{\perp}(j) - \chi_{\perp,b}) = \frac{1}{8J} 2^{-3/2}\eeq
So, at leading order in $1/S$ the edge susceptibility is positive, 
moreover, the approach of $\chi_\perp(j)$ to its bulk value is governed by an oscillating exponential decay. Based on our continuum treatment in the previous section, we expect these results to be strongly modified at higher orders in $1/S$. Indeed, at $T = 0$, from eq. (\ref{chiT0}) on large distances $\chi_\perp(x) - \chi_{\perp,b}$ falls off as $1/x$. However, the coefficient of this power law is of order $1/S$ and, hence, is not captured by the leading order result (\ref{chiS}). When integrated over all space, the large distance power law, which is subleading in the $1/S$ expansion, will lead to a logarithmic divergence in the size/inverse temperature of the system, which would overpower the leading term in $1/S$ coming from short distances. Thus, the combination of eqs. (\ref{chiT0}), (\ref{chiS}) naturally explains the results of Monte Carlo simulations, which see a positive susceptibility of the ``dangling" edge spin combined with the negative total edge susceptibility coming from a large distance tail in $\chi(x)$.


\section{\bf The Comb Structure}
\label{sec:comb}

In this section we explain the appearance of the comb structure (Fig. \ref{figcomb}), seen near the edge in recent Monte Carlo simulations. In our description, we assume the existence of a dynamic valence-bond-solid (VBS) order parameter $V(x)$ with a large correlation length in the N\'eel state. Our treatment becomes exact near a phase transition into a valence-bond-solid phase. This phase transition has attracted a lot of attention in the recent years as it lies outside the Landau-Ginzburg paradigm.\cite{deconfined} It is described by the hedgehog suppressed O(3) $\sigma$-model, with the valence-bond-solid order parameter $V(x)$ being the hedgehog insertion operator. However, the particular details of the phase transition will not be important for our discussion below.

\begin{figure}[t]
\begin{center}
\includegraphics[angle=0,width = 0.4\textwidth]{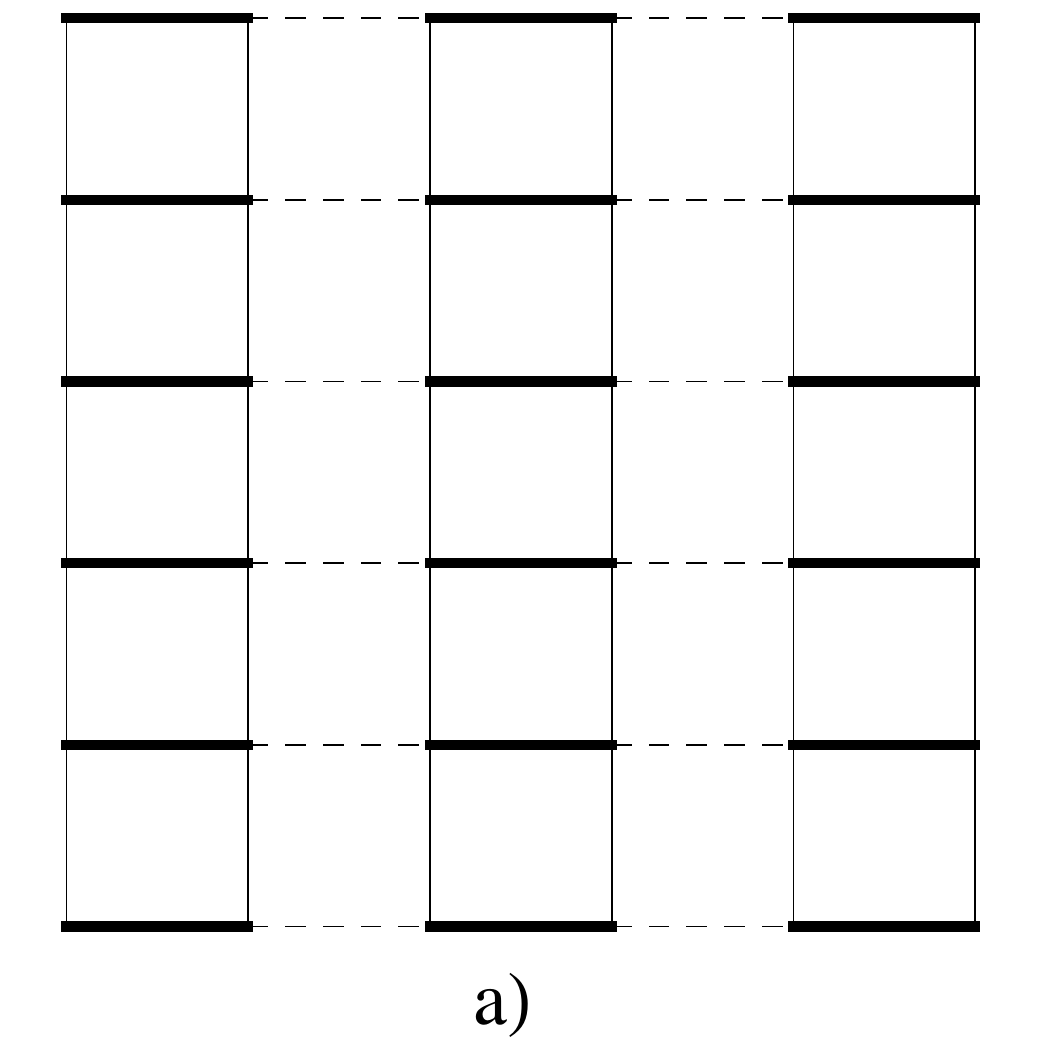}\includegraphics[angle=0,width = 0.4\textwidth]{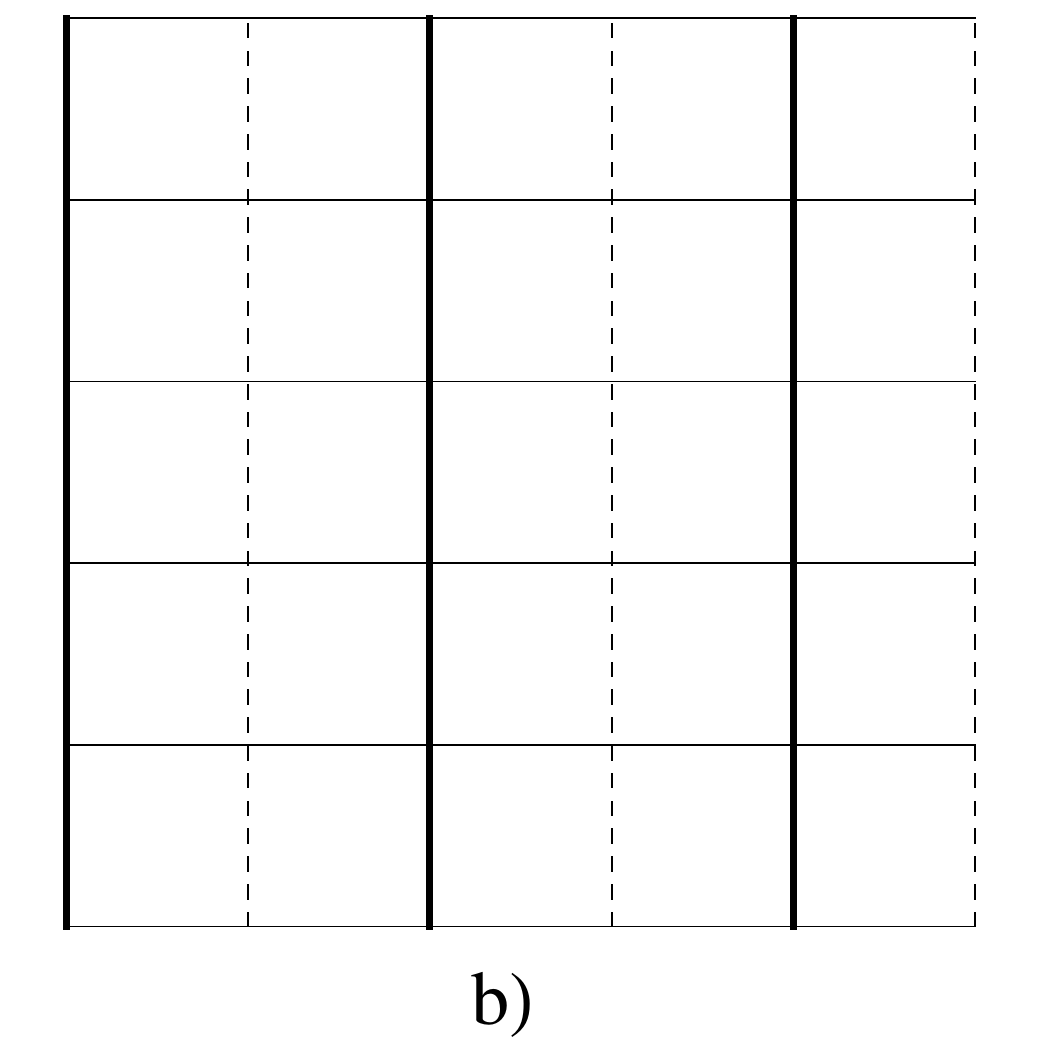}
\caption{a) Lattice order with $\langle V_x\rangle \neq 0$. b) Bond order with $\langle O_x\rangle \neq 0$.}
\label{figorders}
\end{center}
\end{figure}

We begin by defining a microscopic VBS order parameter (which lives on the direct lattice), 
\bea V_x(i) &=& (-1)^{i_x+1/2}\left(\vec{S}(i) \vec{S}(i + \hat{x})-\vec{S}(i) \vec{S}(i - \hat{x})\right)\\
V_y(i) &=& (-1)^{i_y+1/2}\left(\vec{S}(i) \vec{S}(i + \hat{y})-\vec{S}(i) \vec{S}(i - \hat{y})\right)\eea
In this section, we take the origin to lie on the dual lattice. It is customary to group $V_x$, $V_y$ into a complex order parameter $V = V_x + i V_y$ which has the following transformation properties under elements of the square lattice space group:
\bea T^{\dagger}_x V(i_x,i_y) T_x  &=& - V^{\dagger}(i_x-1,i_y)\\
T^{\dagger}_y V(i_x,i_y) T_y &=& V^{\dagger}(i_x,i_y-1)\\
I^{\dagger{\mathrm{dual}}}_{x} V(i_x,i_y) I^{\mathrm{dual}}_{x} &=& V(-i_x,i_y)\\
I^{\dagger{\mathrm{dual}}}_{y} V(i_x,i_y) I^{\mathrm{dual}}_{y} &=& V(i_x,-i_y)\\
R^{\dagger{\mathrm{dual}}}_{\pi/2}V(i_x,i_y) R^{\mathrm{dual}}_{\pi/2} &=& i V^{\dagger}(i_y,-i_x).
\eea
Here $T_{x,y}$ are translations by one lattice spacing in the $x,y$ directions,
$I_{x,y}^{\mathrm{dual}}$ are $x,y$, reflections about a dual lattice point, and $R^{\mathrm{dual}}_{\pi/2}$ is a 90$^\circ$
rotation about about a dual lattice point.
For completeness we also list the transformation property of $V$ under rotations about direct lattice point $(-1/2,-1/2)$,
\beq R^{\dagger{\mathrm{dir}}}_{\pi/2}V(i_x,i_y) R^{\mathrm{dir}}_{\pi/2} = i V(i_y,-1-i_x)\eeq
A non-zero expectation value of the VBS order parameter $V$ would lead to a bond pattern shown in Fig. \ref{figorders} a).
As already noted, the operator $V(x)$ is represented by the hedgehog insertion operator in the continuum description of the antiferromagnet - valence bond solid transition.

Clearly, the order parameter $V$ is adequate for describing the oscillations of horizontal bonds in the comb structure (Fig. \ref{figcomb}).  However, the oscillations of the vertical lines in the comb structure (Fig. \ref{figcomb}), shown separately in Fig. \ref{figorders} b) are not of the ``dimer form." To describe them, we introduce a new order parameter,

\bea O_x(i) &=& (-1)^{i_x} \left(\vec{S}(i+\frac{1}{2}\hat{x} + \frac{1}{2}\hat{y}) \vec{S}(i + \frac{1}{2} \hat{x} -\frac{1}{2} \hat{y})-\vec{S}(i-\frac{1}{2}\hat{x} + \frac{1}{2}\hat{y}) \vec{S}(i - \frac{1}{2} \hat{x} -\frac{1}{2} \hat{y})\right)\nn\\
O_y(i) &=& (-1)^{i_y} \left(\vec{S}(i+\frac{1}{2}\hat{y} + \frac{1}{2} \hat{x}) \vec{S}(i +\frac{1}{2} \hat{y}- \frac{1}{2} \hat{x})-\vec{S}(i - \frac{1}{2}\hat{y}+\frac{1}{2}\hat{x}) \vec{S}(i- \frac{1}{2} \hat{y}- \frac{1}{2} \hat{x})\right)\nn\eea
$O_x$ describes vertical bond lines which are oscillating in strength along the $x$ direction (see Fig. \ref{figorders} b)). Similarly, $O_y$ describes horizontal bond lines, which are oscillating in strength along the $y$ direction. 

We can group $O_x$ and $O_y$ into a single complex order parameter $O = O_x + i O_y$. The transformation properties of $O$ are,
\bea T^{\dagger}_x O(i_x,i_y) T_x &=& - O^{\dagger}(i_x-1,i_y)\\
T^{\dagger}_y O(i_x,i_y) T_y &=& O^{\dagger}(i_x,i_y-1)\\
I^{\dagger{\mathrm{dual}}}_x O(i_x,i_y) I^{\mathrm{dual}}_x &=& -O^{\dagger}(-i_x,i_y)\\
I^{\dagger{\mathrm{dual}}}_y O(i_x,i_y) I^{\mathrm{dual}}_y &=& O^{\dagger}(i_x,-i_y)\\
R^{\dagger{\mathrm{dual}}}_{\pi/2}O(i_x,i_y) R^{\mathrm{dual}}_{\pi/2} &=& i O(i_y,-i_x)\eea
and for rotations about direct lattice point $(-1/2,-1/2)$:
\beq R^{\dagger{\mathrm{dir}}}_{\pi/2}O(i_x,i_y) R^{\mathrm{dir}}_{\pi/2} = i O^{\dagger}(i_y,-1-i_x)\eeq

Now we may ask whether it is possible in the continuum to construct an operator with the transformation properties of $O(x)$ out of $V(x)$.
Clearly, any function of $V$ with no derivatives cannot do the job, since under dual lattice reflections $I^{\mathrm{dual}}_{x,y}$, $O$ transforms non-trivially, while $V$ transforms trivially. Thus, a static uniform condensate of $V$ (not surprisingly) cannot give rise to the order in Fig. \ref{figorders} b).
However, we can obtain an expression with the transformation properties of $O$ if we allow for derivatives of $V$. Considering expressions with one power of $V$ and one derivative, we obtain,
\beq O_x \sim \d_x V_x, \quad O_y \sim \d_y V_y \label{OVconn}\eeq
(with the same proportionality constant).

Thus, if dimerization of horizontal bonds is present and is inhomogeneous along the $x$ direction then we automatically obtain the ``secondary" order in Fig. \ref{figorders} b). 

Now, we may ask, how a non-zero expectation value of the VBS order is generated? Indeed, in the N\'eel phase, in the bulk, the $\mathbb Z_4$ lattice rotation symmetry is unbroken and $\langle V \rangle = 0$. However, the edge possesses a smaller lattice symmetry group than the bulk - in particular, the lattice rotation symmetry is explicitly broken. This is manifested in the continuum formulation by the appearance of an edge perturbation,
\beq \delta S = \frac{1}{2} h \int d\tau dy \,(V + V^{\dagger})  =  h \int d\tau dy \,V_x
\label{edgeaction} 
\eeq
In the phase where $V$ is gapped, we expect such a coupling will lead to an appearance of $\langle V_x(x,y) \rangle$ decaying away from the edge. Hence, we will also have $\langle O_x(x,y) \rangle \neq 0$, which close to the critical point can just be obtained from (\ref{OVconn}). Thus, the appearance of the comb structure is very natural. 

Based on the known results on boundary critical behaviour,\cite{diehl} we may write down the scaling forms for $\langle V(x)\rangle$, $\langle O(x)\rangle$ in the critical region. 
The edge perturbation $\delta S$ is relevant at the critical point provided that $\Delta^V < 2$, where $\Delta^V$ is the scaling dimension of operator $V(x)$. Then the scaling forms become universal (up to overall multiplicative factors),
\bea \langle V_x(x) \rangle &\sim& \frac{1}{\xi^{\Delta^V}} g(x/\xi)\label{Vscal}\\
\langle O_x(x) \rangle &\sim& \frac{1}{\xi^{\Delta^V +1}} g'(x/\xi)\eea
Here $\xi$ is the correlation length of the VBS order parameter in the Neel phase (which is proportional to the inverse spin stiffness $c/\rho_s$ with some universal amplitude). In the deconfined criticality scenario, $\xi$ will be given by the inverse skyrmion mass. Note that due to the extra derivative in $O$ compared to $V$, the modulations of lines parallel to the edge become parametrically weaker than those of dimers perpendicular to the edge as we approach the phase transition. We may also write down short and long distance asymptotics of $g(u)$,
\bea g(u) &\sim& \frac{1}{u^{\Delta^V}}, \quad u \to 0\label{Vscalshort}\\
g(u) &\sim& e^{-u}, \quad u \to \infty\label{glarge}\eea
where we have not specified the likely power-law prefactor for the long distance asymptotic (\ref{glarge}).



\section{A non-magnetic impurity}
\label{sec:imp}

This section will briefly discuss the case of a different defect in a perfect square lattice antiferromagnet: a single
site with a missing spin. This is often experimentally realized in Cu antiferromagnets by replacing Cu with Zn.
We are interested in the configuration of VBS order around this impurity---this was addressed recently in 
Ref.~\onlinecite{kmms} using methods similar to those used in Section~\ref{sec:comb}. Our purpose is to 
connect these phenomenological approaches to the field-theoretic treatment near the deconfined critical
point presented in Ref.~\onlinecite{max2}.

As in Section~\ref{sec:comb}, we begin by describing the influence of the impurity by writing down the action
for $V$ consistent with the symmetries of the impurity Hamiltonian; here the action has to be invariant under
$R_{\pi/2}^{\rm direct}$, $I_{x}^{\rm direct}$, and time-reversal. Then the analog of the edge perturbation in Eq.~(\ref{edgeaction})
for an impurity at $\vec{x}_{\mathrm{imp}}$ is\cite{kmms} 
\begin{equation}
{S}_{{\rm imp},V} = - \lambda_1 \int d \tau \left. \left( \frac{\partial V}{\partial x} + \frac{\partial V^\dagger}{\partial x}
+ i \frac{\partial V}{\partial y}  - i \frac{\partial V^\dagger}{\partial y} \right) \right|_{\vec{x}=\vec{x}_{\mathrm{imp}}}
\label{simpv}
\end{equation}
As shown in Ref.~\onlinecite{kmms}, this perturbation induces `vortices' in the VBS order around the impurity.
We now discuss the origins of the term
 $\mathcal{S}_{{\rm imp},V}$ in the critical theory of the N\'eel-VBS transition in the insulator. This will determine the
behavior of the coupling $\lambda_1$ near this transition.

The behavior of a non-magnetic impurity near this transition has been described in Refs.~\onlinecite{kolezhuk,max1,max2}.
For the bulk model without an impurity, a field theoretic description of the vicinity of the quantum critical point \cite{deconfined,SachdevMurthy,mv} is provided by the $\mathbb{CP}^{N-1}$ theory at $N=2$:
\begin{equation}
\mathcal{S} = \int d^2 r d \tau \left[ |(\partial_\mu - i A_\mu)z_\alpha |^2 + s |z_\alpha|^2 + \frac{g}{2} \left( |z_\alpha|^2 \right)^2 + \frac{1}{2e^2} (\epsilon_{\mu\nu\lambda} \partial_\nu A_\lambda)^2 \right]. \label{cpn}
\end{equation}
Here $\mu,\nu,\lambda$ are spacetime indices, 
$z_\alpha$, $\alpha = 1 \ldots N=2$ is a complex scalar which is a SU($N$) fundamental, and $A_\mu$ is a non-compact U(1) gauge field.
As discussed in Ref.~\onlinecite{kolezhuk}, the most important perturbation to Eq.~(\ref{cpn}) induced by 
the non-magnetic impurity near the deconfined critical point is the impurity Berry phase:
\begin{equation}
\mathcal{S}_{\rm imp} = i Q \int d\tau A_\tau (\vec{x} = 0, \tau) \label{simp}
\end{equation}
where $Q$ is a `charge' characterizing the impurity. The value of $Q$ does not flow under the RG,
and so $Q$ is a pure number which controls all universal characteristics of the impurity response. 

Let us now discuss the symmetries of $\mathcal{S} + \mathcal{S}_{\rm imp}$.
In addition to the global SU($N$) symmetry, this model has a global U(1)$_\varphi$  symmetry which is the dual of the 
U(1) gauge invariance. The primary action of this symmetry is on the monopole operator, $V (\vec{x} , \tau)$, which 
transforms as
\begin{eqnarray}
{\rm U}(1)_{\varphi} &:& V \rightarrow V e^{i \varphi} \nonumber \\
R_\theta &:& V \rightarrow V
\label{vmpsg}
\end{eqnarray}
At the moment, this U(1)$_\varphi$ `flux' symmetry is independent of spatial rotations $R_{\theta}$, and this has been indicated
above for completeness. The physical $\mathbb{Z}_4$ lattice rotation symmetry is the combination of $\pi/2$ rotations in $U(1)_\varphi$ and $R_\theta$ - thus, the monopole operator $V$ is identified with the VBS order parameter.

A key property \cite{max2} of the theory $\mathcal{S}+\mathcal{S}_{\rm imp}$
is the operator product expansion for the monopole operator $V$ in the vicinity of the impurity
\begin{equation}
\lim_{|\vec{x}| \rightarrow 0} V (\vec{x}, \tau) \sim  |\vec{x}|^{\Delta^V_{\rm imp}} \, e^{-i Q \theta} \, V_{\rm imp} (\tau)
\label{vimp5}
\end{equation}
where $\theta$ is the azimuthal angle of $\vec{x}$, and $\Delta^V_{\rm imp}$ is the impurity correction to the 
scaling dimension of $V$ ($\Delta^V$) as defined in Ref.~\onlinecite{max2}.
Here $V_{\rm imp}$ is a fluctuating impurity degree of freedom with a non-trivial scaling dimension. The presence of the $e^{-iQ \theta}$
factor  indicates a $Q$-fold winding in the phase of the VBS order parameter around the impurity. Thus, the effect of the impurity Berry phase term is to
induce vortex-like correlations in bond order near the impurity. 

However, the way this vortex is pinned to the lattice is determined by additional impurity perturbations, the most relevant of which is given by Eq. (\ref{simpv}). We can understand this by continuing our symmetry analysis.
The combination of Eqs.~(\ref{vmpsg}) and (\ref{vimp5}) implies the following transformations of $V_{\rm imp}$ under
the flux symmetry and spatial rotations:
\begin{eqnarray}
{\rm U}(1)_{\varphi} &:& V_{\rm imp} \rightarrow V_{\rm imp} e^{i \varphi} \nonumber \\
R_\theta &:& V_{\rm imp} \rightarrow V_{\rm imp} e^{-i \theta} 
\label{vimppsg}
\end{eqnarray}
Here, and henceforth, we specialize to the case $Q=1$, although the generalization to other $Q$ is not difficult.  We note that the
quantum numbers of $V_{\mathrm{imp}}$ are the same as those of the perturbation (\ref{simpv}). Hence, the two will mix and we may replace (\ref{simpv}) by,
\beq {S}'_{\mathrm{imp},V}= - \lambda'_1 \int d \tau\, V_{\mathrm{imp}}(\tau)\label{simpv2}\eeq
Now, there are two possibilities. If the perturbation (\ref{simpv2}) is relevant at the critical point, which occurs for 
\beq \mathrm{dim}[V_{\rm{imp}}] = \Delta^V + \Delta^V_{\rm imp}<1\label{relevant}\eeq
the coupling $\lambda'_1$ will flow to infinity. In this case, at criticality, the VBS order parameter will be given by,
\beq \langle V(\vec{x},\tau) \rangle \sim \frac{ e^{i \theta}}{|\vec{x}|^{\Delta^V}}\eeq
Alternatively, if the coupling $\lambda'_1$ is irrelevant, we can treat it in perturbation theory and obtain,
\beq \langle V(\vec{x},\tau) \rangle \sim  \frac{ e^{i \theta}}{|\vec{x}|^{2 \Delta^V + \Delta^V_{\rm{imp} - 1}}}.\eeq

Now let us move into the Coulomb phase of $\mathcal{S}$, where there is a mass gap, $m$, for the $z_\alpha$ spinons.
We are interested in the effective theory for $V(x)$ at energy scales smaller than this
mass gap. The only low energy degree of freedom is the (pseudo)-Goldstone $\varphi$ associated with spontaneous breaking of the $U(1)_\varphi$ symmetry.
We identify, $V \sim m^{\Delta^V} e^{i \varphi}$. The effective action for the $\varphi$ field in the absence of impurity takes the form,
\beq {S} = \int d^2 x d \tau\, \left(\frac{e^2}{2 (2 \pi)^2} (\d_{\mu} \varphi)^2  - \lambda_4 \cos(4 \varphi)\right)\label{Leff}\eeq
Here $e^2 \sim m$ is the effective electric charge in the Coulomb phase and $\lambda_4 \sim m^{{\mathrm{dim}}[V^4]}$ is the dangerously irrelevant perturbation that breaks $U(1)_\varphi$ symmetry to the physical ${\mathbb Z}_4$.
Now, let's discuss the impurity perturbations in the effective theory. One such perturbation can be simply obtained from (\ref{simpv}) by replacing $V \to e^{i \varphi}$,
\beq S_{\mathrm{imp},\mathrm{eff}} = - \lambda_{1,\mathrm{eff}} \int d\tau \left( i (\d_x \varphi + i \d_y \varphi) e^{i \varphi} + h.c.\right)\eeq 
We are interested in how the coefficient of this term $\lambda_{1,\mathrm{eff}}$ is renormalized. If the perturbation (\ref{simpv}) is relevant (see eq. (\ref{relevant})), the impurity response will be universal and $\lambda_{1, \mathrm{eff}}$ will be a constant in $m$ by dimensional analysis. Otherwise, $\lambda_{1, \mathrm{eff}} \sim m^{\Delta^V + \Delta^V_{\rm{imp}} - 1}$. 





\section{\bf Conclusion}
\label{sec:concl}
In this paper we have addressed two puzzles raised by recent Monte Carlo simulations of edge response in square lattice quantum antiferromagnets. The first puzzle is the appearance of negative edge susceptibility - we have shown that this effect is due to low energy spin-waves. We predicted that the total edge susceptibility diverges logarithmically as inverse temperature/system size goes to infinity, and found
this to be in good agreement with the Monte Carlo simulations of Ref.~\onlinecite{Hoglund}.  
We would like to note here that our results on the low temperature behaviour of susceptibility apply equally well to a clean and rough edge, as our continuum O(3) $\sigma$-model description does not assume translational invariance along the edge. (However, for the rough edge, there may be additional important contributions to the 
susceptibility coming from Berry phase effects, not present in the O(3) $\sigma$ model.) The second puzzle is the observation of a comb structure in the bond response near the edge. We have argued that this is likely a purely quantum mechanical effect, which cannot be captured by the naive $1/S$ expansion. We have shown that the appearance of the comb structure can be understood in the framework of a continuum theory involving a dynamical valence-bond-solid order parameter. Such a description becomes exact in the neighbourhood of a quantum phase transition to a valence-bond-solid phase. We hope that the simulations of edge response in Heisenberg model\cite{Hoglund} will be extended to the so-called $JQ$ model where such a phase transition is observed.\cite{SandvikVBS,MelkoKaul} We have made a few predictions regarding the behaviour of the comb structure near criticality, e.g. the relation between the behaviour of bonds parallel and perpendicular to the edge in the comb. Edge response near the quantum critical point might also be a viable way to extract the scaling dimension of the valence-bond-solid order parameter, see eqs. (\ref{Vscal}),(\ref{Vscalshort}). 

Finally, in Section~\ref{sec:imp}, we briefly discussed some related issues on the problem of a single
non-magnetic impurity, complementary to the more detailed discussion of this case in Ref.~\onlinecite{kmms}.

\acknowledgements
We are very grateful to Anders Sandvik and Kaj H\"oglund for informing us about their results prior to
publication and for useful discussions. This research was supported by the NSF under grant DMR-0757145.

\appendix
\section{Eigenfunctions of Bogoliubov quasiparticles}\label{app:eigen}
First, we define for fixed energy $\lambda$,
\beq \left(\begin{array}{c} u\\v\end{array}\right) = \frac{1}{\sqrt{2 \lambda}}\left(\begin{array}{c} - \sqrt{1+\lambda}\\\sqrt{1- \lambda}\end{array}\right)\eeq
Now, the eigenstates can be expressed as,

\vspace{0.5cm}
Normalizable solution:
\bea
\lambda &=& \frac{1}{\sqrt{2}} |\sin k_y|, \quad -\pi/2 < k_y < \pi/2\\
\phi(j) &=& c_1\left(\begin{array}{c}u\\v\end{array}\right) e^{-s_1 j} + c_2 \left(\begin{array}{c}-u\\v\end{array}\right) (-1)^j e^{-s_2 j}\\
e^{s_1} &=& (\sqrt{2}+1)(\sqrt{1+\cos^2 k_y} - \cos k_y), \quad e^{s_2} = (\sqrt{2}+1)(\sqrt{1+\cos^2 k_y} + \cos k_y)\nn\\
\left(\begin{array}{c} c_1\\c_2 \end{array}\right) &=& \frac{2^{-\frac34}(\sqrt{2}-1) |\sin k_y|}{\sqrt{1-|\sin k_y|}\sqrt{1+\cos^2 k_y}}\left(\begin{array}{c} e^{s_2} \sqrt{1-\lambda} - \sqrt{1+\lambda}\\e^{s_1} \sqrt{1-\lambda} - \sqrt{1+\lambda}\end{array}\right)\nn\\
\eea

Continuum\, solutions:
\bea
\gamma &=& \frac{1}{2} (\cos(k_x) + \cos(k_y)), \quad \lambda = \sqrt{1 - \gamma^2}, \quad 0< k_x < \pi-|k_y|, \quad -\pi/2 < k_y < \pi/2\nn\\\label{econt}\eea

Branch 1: \quad $0 < k_x < \cos^{-1}(1-2 \cos{k_y})$
\bea
\phi(j) &=& \frac{1}{|\alpha|}\left(\begin{array}{c} (\alpha e^{i k_x j} + \alpha^{*} e^{-i k_x j} - (-1)^j e^{-j s}) u \\
(\alpha e^{i k_x j} + \alpha^{*} e^{-i k_x j} + (-1)^j e^{-j s})v\end{array}\right), \quad s = \cosh^{-1}(\cos k_x + 2 \cos k_y)\\
\alpha &=& - \frac{1}{2 \lambda}\left( \gamma e^s -1 - \frac{i}{\sin k_x} ((\gamma \cos k_x - 1) e^s + \gamma - \cos k_x)\right) \nn\\\eea

Branch 2: \quad $\cos^{-1}(1-2 \cos{k_y}) < k_x < \pi - |k_y|$
\bea \tilde{k}_x &=& \pi - \cos^{-1}(\cos(k_x) + 2 \cos(k_y)), \quad \pi - |k_y| < \tilde{k}_x < \pi\\
\phi_1 (j) &=& A
\left( c_{11} \cos(k_x (j+1/2)) \left(\begin{array}{c}
u \\ v\end{array}\right) + c_{12}  \cos(\tilde{k}_x (j+1/2)) \left(\begin{array}{c}
-u \\ v\end{array}\right)\right)\\
\phi_2 (j) &=& A
\left( c_{21} \sin(k_x (j+1/2)) \left(\begin{array}{c}
u \\ v\end{array}\right) + c_{22}  \sin(\tilde{k}_x (j+1/2)) \left(\begin{array}{c}
-u \\ v\end{array}\right)\right)\\
A &=& (\sin k_x)^{\frac12} (\sin((k_x + \tilde{k}_x)/2) + \gamma \sin((k_x - \tilde{k}_x)/2))^{-\frac12}\\
c_{1 1} &=& (1+ \gamma)^{\frac12} \left(\frac{2 \cos \tilde{k}_x/2}{\cos k_x/2}\right)^{\frac12},\quad c_{12} = 
(1- \gamma)^{\frac12} \left(\frac{2 \cos k_x/2}{\cos \tilde{k}_x/2}\right)^{\frac12}\\
c_{2 1} &=& (1- \gamma)^{\frac12} \left(\frac{2 \sin \tilde{k}_x/2}{\sin k_x/2}\right)^{\frac12},\quad c_{22} = 
(1+ \gamma)^{\frac12} \left(\frac{2 \sin k_x/2}{\sin \tilde{k}_x/2}\right)^{\frac12}
\eea
The division of the continuum spectrum into two branches is clear when we look at a plot of $\lambda(k_x)$ (\ref{econt}) for $k_y$ fixed.  For $\sqrt{1-\cos^4(k_y/2)} < \lambda < \sqrt{1-\sin^4(k_y/2)}$ there is only one corresponding value of $k_x$ in the range $0 < k_x < \pi$ (there is always a solution with opposite $k_x$, as well). This is our branch 1. On the other hand, for $\sqrt{1-\sin^4(k_y/2)} < \lambda < 1$ there are two solutions with $0 < k_x < \pi$, which we label by $k_x$ and $\tilde{k}_x$. These two solutions are mixed by the edge and form the two linearly independent eigenstates $\phi_1$, $\phi_2$ in branch 2.

\end{document}